# Global city densities: re-examining urban scaling theory

Joseph R. Burger[1,*], Jordan G. Okie[2], Ian Hatton[3], Vanessa P. Weinberger[4], Munik Shrestha[5], Kyra J. Liedtke[1], Tam Be[6], Austin R. Cruz[7], Xiao Feng[8], Cesar Hinojo-Hinojo[9], Abu S.M.G. Kibria[10], Kacey C. Ernst[11], Brian J. Enquist[7,12]

[1] *Department of Biology, University of Kentucky, Lexington, KY 40502, USA: https://orcid.org/0000-0002-7361-3858;* *Corresponding author: robbieburger@uky.edu
[2] *School of Earth and Space Exploration, Arizona State University, Tempe, AZ 85287, USA*
[3] *Max Plank Institute for Mathematics in the Sciences, Leipzig, 04103, Germany*
[4] *GEMA Center for Genomics, Ecology & Environment, Universidad Mayor, Santiago, Chile*
[5] *Network Science Institute, Northeastern University, Boston, MA, 02115, USA*
[6] *Department of Computer Science, University of Arizona, Tucson, AZ 85721, USA*
[7] *Department of Ecology and Evolutionary Biology, University of Arizona, Tucson, AZ 85721, USA*
[8] *Department of Geography, Florida State University, Tallahassee, FL 32306-2190, USA*
[9] *School of Geographical Sciences and Urban Planning, Arizona State University, Tempe, AZ 85281, USA*
[10] *School of Natural Resources and the Environment, University of Arizona, Tucson, AZ 85721, USA*
[11] *Department of Epidemiology & Biostatistics, University of Arizona, Tucson, AZ 85721, USA*
[12] *Santa Fe Institute, Santa Fe, NM 87501, USA*

**Abstract:** Understanding scaling relations of social and environmental attributes of urban systems is necessary for effectively managing cities. Urban scaling theory (UST) has assumed that population density scales positively with city size. We present a new global analysis using a publicly available database of 933 cities from 38 countries. Our results showed that (18/38) 47% of countries analyzed supported increasing density scaling (pop ~ area) with exponents ~⅚ as UST predicts. In contrast, 17 of 38 countries (~45%) exhibited density scalings statistically indistinguishable from constant population densities across cities of varying sizes. These results were generally consistent in years spanning four decades from 1975 to 2015. Importantly, density varies by an order of magnitude between regions and countries and decreases in more developed economies. Our results (i) point to how economic and regional differences may affect the scaling of density with city size and (ii) show how understanding country- and region-specific strategies could inform effective management of urban systems for biodiversity, public health, conservation and resiliency from local to global scales.

**200 word statement of contribution:** Urban Scaling Theory (UST) is a general scaling framework that makes quantitative predictions for how many urban attributes spanning physical, biological and social dimensions scale with city size; thus, UST has great implications in guiding future city developments**.** A major assumption of UST is that larger cities become denser. We evaluated this assumption using a publicly available global dataset of 933 cities in 38 countries. Our scaling analysis of population size and area of cities revealed that while many countries analyzed showed increasing densities with city size, about 45% of countries showed constant densities across cities. These results question a key assumption of UST. Our results suggest policies and management strategies for biodiversity conservation, public health and sustainability of urban systems may need to be tailored to national and regional scaling relations to be effective.

## Introduction

Urbanization has significantly increased in recent decades. Developing a science of cities is necessary to understand, predict and manage the future of an urbanizing planet (Acuto et al., 2018; Burger et al., 2019; Bettencourt, 2021; Uchida et al., 2021). A general theory of cities should predict how variation in city size, infrastructure, and governance are linked to influence biodiversity, urban ecosystems, resource use, social interactions, innovations, economic activity, public health, crime, and urban sustainability (Bettencourt et al., 2007; Lobo et al., 2019; Bettencourt, 2020; 2021; Uchida et al., 2021). Indeed, while many socioeconomic, behavioral, ecological, and evolutionary processes are affected by urbanization, it is unclear how city size impacts these processes directly or indirectly via variation in human population density.

Several studies have recently extended scaling approaches from biology一where various organismal characteristics scale with size一to the study of cities (Bettencourt et al., 2007; Bettencourt, 2013, 2020; West, 2018; Lobo et al., 2019; Ortmann et al., 2020). Urban Scaling Theory (UST; a.k.a Urban Settlement Theory) is a general scaling framework that makes quantitative predictions for many urban attributes spanning physical, biological, and social dimensions, and how they scale with city size. Similar to the scaling of traits with body size in biology, various attributes of cities scale as a power-law:

$$Y(t) = Y_0\ (t)\ N(t)^{\alpha} \qquad \text{(Eqn. 1)}$$

where $Y$ is an attribute for a given city at time $(t)$, $N$ is human population size or total number of inhabitants in the city, $Y_0$ is the elevation of the normalization constant and $\alpha$ is the exponent. The three main classes of scaling behavior are: (i) *Sublinear* ($\alpha$ > 1)*, (ii) linear* ($\alpha$ = 1, aka isometric), and (iii) *superlinear scaling* ($\alpha$ > 1). Analysis of initial data from modern cities in a few countries suggest $\alpha$ < 1 for infrastructural quantities; $\alpha$ = 1 for measures of resource use and waste production; and $\alpha$ > 1 for social attributes including super-creative employment, crime, and infectious disease (Bettencourt et al., 2007). These empirical patterns have helped develop the mechanistic theory behind the scaling categories, with spatial filling networks and social interaction dynamics in growing urban systems as their determinants (Bettencourt, 2013).

A key underlying assumption of UST mechanistic theory is that densities increase with city size, which results in *economies of scale* ($\alpha$ < 1) in space use (Bettencourt 2013) that influences the scaling of several characteristics of cities, particularly those related to social outcomes. This results in exponents of $\alpha$ > 1 for some urban features are due to efficiencies gained in communication and infrastructure networks in larger cities. More specifically, assuming optimal infrastructure and material use, UST proposes that circumscribing urban areas and the urban built area scale with population size as $\alpha = 2/3$ and $\alpha = 5/6$, respectively. However, it is essential to establish empirical foundations to advance general theory and applications for global urban management practices.

Population density is also a critical variable of study for other areas of fundamental urban science and practical purposes. Variation in human population density reflects the unique ecology of hyper-dense urban societies existing at several orders of magnitude denser compared to hunter-gather and early agrarian societies (Burger et al., 2017). Density fundamentally affects social interactions, the built environment, open/green spaces, demand for natural resources and impacts on landscapes. The null expectation is that population density is invariant of city size if city area scales isometrically with population size. However, UST predicts that city area scales sublinearly with city population size and, therefore, human population density increases with city size. Increases in population density suggests important underlying interactions and dynamics and has fundamental implications for the emergent behavior of many urban social attributes.

Despite the apparent importance of understanding variation in urban densities, there is limited understanding, however, of how the scaling of area with population size vary globally. Some studies have supported sublinear scaling of areas with population size based on ordinary least-square regression and limited data that is highly unrepresentative of cities especially in the tropics and "Global South". For example, one seminal study was based on data for just four highly-developed nations: USA, Sweden, United Kingdom, and Canada (Bettencourt 2013). Another study was based on historical societies with small city sizes compared to modern cities (Lobo et al. 2019). A recent global study of the scaling of area with population size used The Organisation for Economic Co-operation and Development (OECD) data for Functional Urban Areas by Ortman et al. (2020) showed mixed results where some countries and regions scaled linearly while others scaled sublinearly at 5/6. So, advancing urban science and UST necessitates assessing if urban area scales sublinearly with a ⅔ or ⅚ exponent. Doing so is the first step in developing a general scaling framework for managing urban biodiversity and sustainable cities.

Here, we examine the scaling behavior of city areas with population size. We test the assumption of increasing densities, people per area, and city size by using a publicly available global database that fills important gaps in urban scaling knowledge (Acuto et al. 2018), especially in tropical regions. In addition, we explored whether factors such as regional differences, economic development status, and time influence the value of the scaling relations of the log population vs log area relationship. We end by discussing the important implications of these findings for managing urban ecosystems and biodiversity at local, regional and global scales.

**Materials and Methods**

*The data*

We assessed the scaling of urban area and population size for 933 cities in 38 countries globally. We used the publicly available Global Human Settlement (GHS) Urban Center Database (https://ghsl.jrc.ec.europa.eu/ghs_stat_ucdb2015mt_r2019a.php; Pesaresi et al. 2019) and the associated Global Human Settlement Layer (GHSL) project (https://ghsl.jrc.ec.europa.eu/index.php; Florczyk et al. 2019). The GHS classifies the physical extent of human settlements – from large megacities to villages and towns using a consistent methodology across the globe with satellite remote sensing as a primary source of information. Supervised automatic data classification is used to build the GHS built-up area grid (GHS-BUILT) using open decametric-resolution satellite imagery collected by the Landsat and Sentinel missions. Using census data through spatial modeling techniques combined with GHS-BUILT, the GHSL project produces global population density grids (GHS-POP). The data integration process foresees the downscaling of the information from the national census district level to a regular, finer-scaled, gridded, built-up density information layer at the resolution of 250m$^2$ and 1 km$^2$. Importantly, the GHS does not assign cities and populations by political boundaries (e.g., MSAs, countries, counties), but rather by the clustering and proximity of the built-up environment classified as urban.

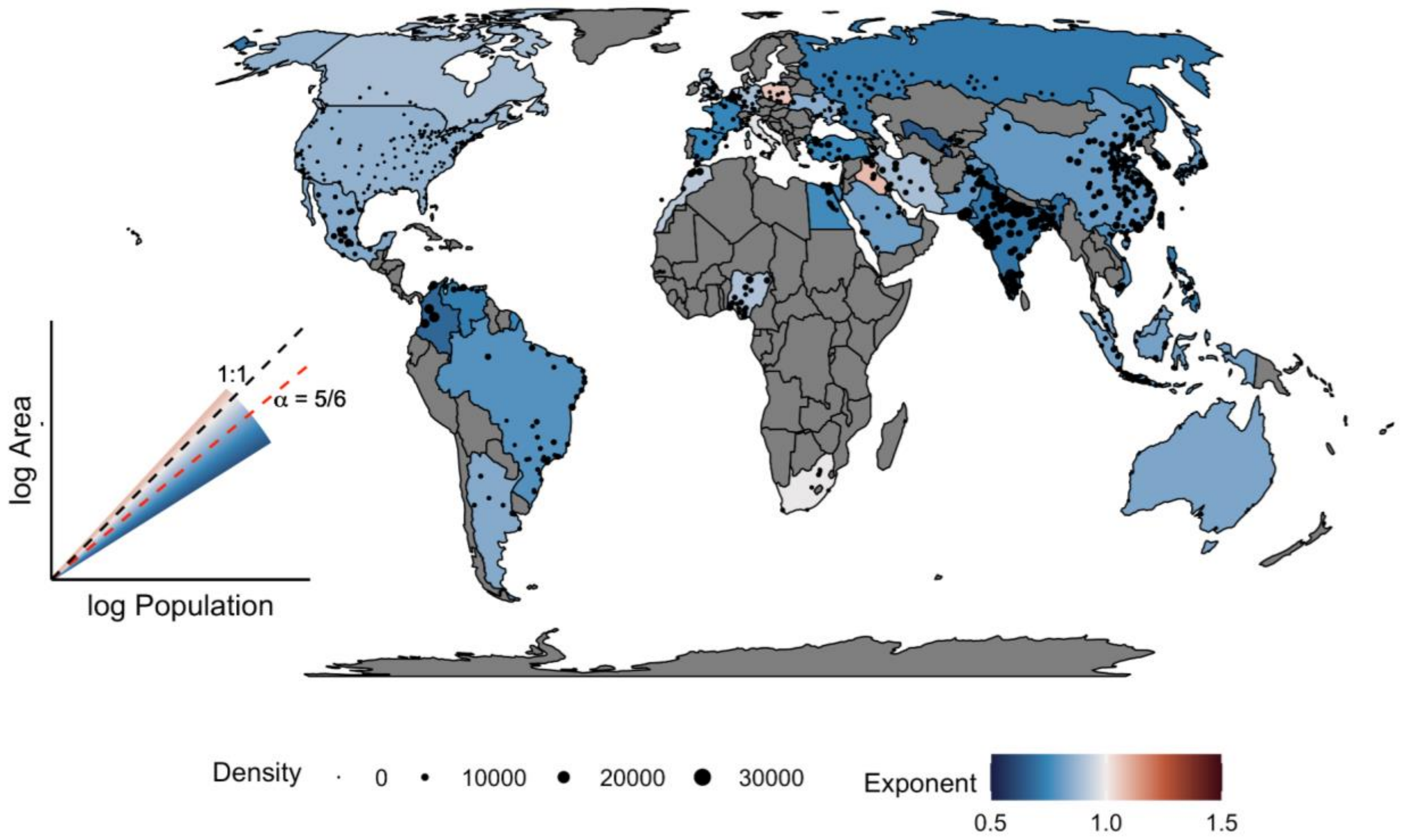

**Figure 1. Global map of city scaling exponents for 38 countries for year 2015 from the GHS. See appendix for scaling statistics by country. Note that colors represent exponents for countries that contain more than 5 cities with an area >100km$^2$ and >50,000 population. Gray countries do not satisfy these criteria; point size represents city density, which varies from <1,000 to >30,000. (see *Appendix 1 for statistics*).**

The combination of GHS-BUILT and GHS-POP produces the GHS settlement model (GHS-SMOD), building on the "Degree of Urbanization" concept (Dijkstra and Poelman, 2014). Distinguishing between three main typologies of human settlements based on population density cut-off values: Urban Centers, Urban Clusters, and Rural Settlements. The Urban Centers, which represent the most densely inhabited part of human settlements, are defined as "the spatially-generalized high-density clusters of contiguous grid cells of 1 km$^2$ with a density of at least 1,500 inhabitants per km$^2$ of land surface or at least 50% built-up surface share per km$^2$ of the land surface, and a minimum population of 50,000", and are the core of the GHS-UCDB dataset. It considers all "Urban Centers" data from the GHS-SMOD 2015, 1 x 1 km$^2$ resolutions, leading to a database containing over 10,000 individual cities and characterizing each by a number of variables describing the geography and the environment of the place, as well as socio-economic parameters and the potential exposure of an Urban Center to natural disasters. Initial analysis of all the data produced erroneous results due to variation in the resolution of the combined datasets in GHS. Further inspection of 'outlier' cities by our international team of scientists further identified problems with small area cities in the dataset in Bangladesh, Nepal, and other countries. The GHS was notified and agreed this was problematic and should be corrected in future versions. So, we excluded cities with total urban areas of <100 $km^2$, reducing the number in our analysis to 933 cities from countries with at least five cities with >50,000 people, in contrast to Ortam et al. (2020) who used

*Statistical Analysis*

We conducted both Reduced Major Axis (RMA) and Ordinary Least Squares (OLS) regression of log population size versus log area to estimate the scaling exponent ($\alpha$) and elevation ( $Y_0$) for countries by years available in the GHS database (yrs. 1975, 1990, 2000, 2015) following Fuller and Gaston (2009) and Gudipudi et al. (2019). We report OLS and RMA regression statistics and scaling parameters in Appendix Table 1. We used the 95% confidence intervals (CIs) of scaling exponents to evaluate whether scaling relations at different scales of analysis (within-in country) are statistically indistinguishable from linear scaling (CIs bounding 1) versus providing support for allometric scaling (CIs not containing 1), including overlapping with $\alpha = 5/6$. The general nature of the results were similar between RMA and OLS, so we report only RMA analyses in the text because it is the standard method for allometric scaling studies ( see Appendix and Supplemental Material for more information). We assessed the effect of the development status of the countries over scaling exponents and elevation through an analysis of covariance (ANCOVA), where log(area) was the dependent variable and population and development status of the country and their interaction were the predictor variables.

## Results

*Country-specific scalings*

Our analysis revealed that the scaling exponents ($\alpha$) varied widely across countries (Figure 1 & 2), from 0.63-to 1.08 (Appendix 1); 17 countries (47%) supported UST predictions of a sublinear scaling exponent of ⅚ for built urban areas. In contrast, 18 countries (48%) had scaling relations indistinguishable from linear ($\alpha = 1$) including 8 countries with 10 or more cities of similar densities (Table 1). See Appendix for country-level statistics for all years.

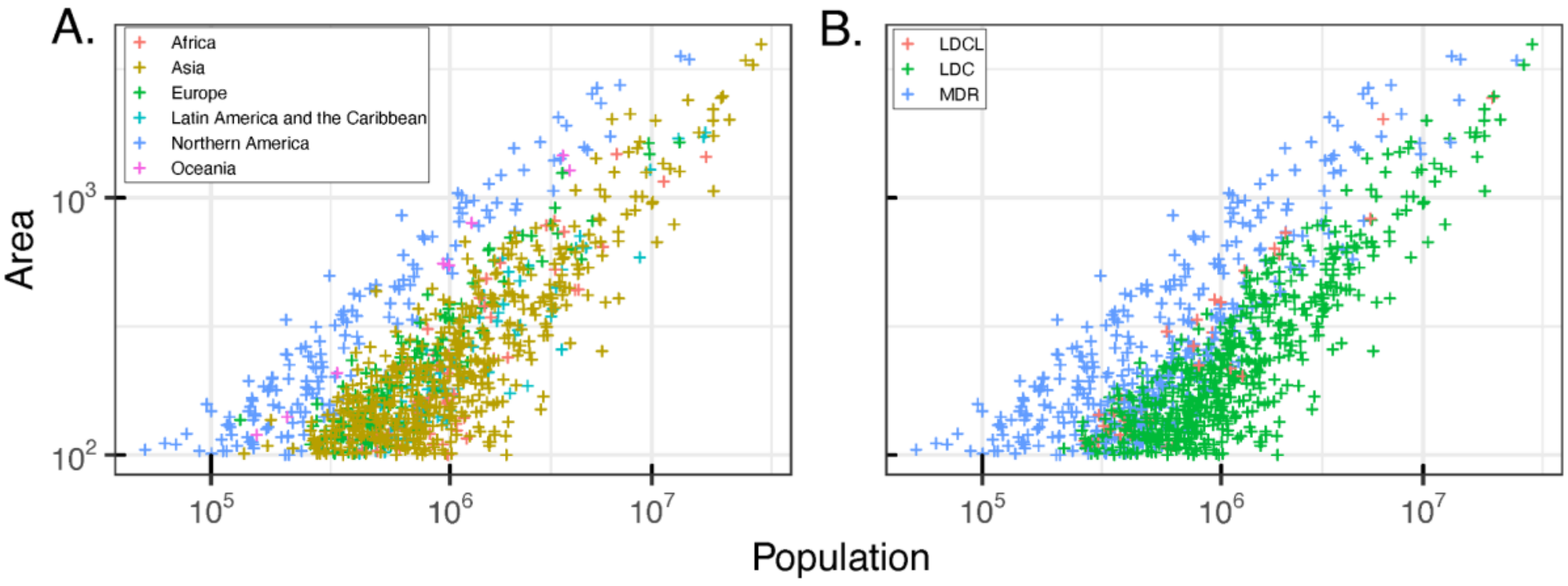

**Figure 2. A scatter plot of 933 cities in 2015 used in this study colored by geographic region (A) and level of economic development status: LDC = low development countries, LDCL = least developed countries, and MDR = more developed countries (B). Note that the x and y axes are log10 transformed.**

We found greater variation in the normalization constant $Y_0$ between countries than within (Appendix 1). Development status had no detectable effect on scaling exponents (F = 1.929, p = 0.146), but affected $Y_0$ (F = 311.254, p < 0.001), being highest in the More Developed countries and lower in the Less and Least Developed countries. Those results indicate that cities in less and least developed countries are denser than the more developed countries for any given city size. For example, the highest density cities in the curated GHS dataset occur in countries of Asia (Surat = 24,117 ind $km^{-2}$), Africa (Cairo = 12,451 ind $km^{-2}$) and tropical Latin America (Salvador = 9,839 ind $km^{-2}$) and are considerably higher for a given city size than densities in the USA (NYC = 2,963 ind $km^{-2}$), Australia (Sydney = 2,762 ind $km^{-2}$), Canada (Toronto = 2,990 ind $km^{-2}$). Maximum densities for each country range from ~2,000-20,000 ind $km^{-2}$, and tend to be higher close to the tropics (Figure 1). See appendix for full statistics by country and years.

*Temporal analysis*

We found scaling exponents to vary over time for some countries and regions and remain invariant in others (Figs 3, and 4; Appendix). Notably, the 95% CIs for the exponents of 13 countries overlap with one (linear) over three decades (Belgium, Colombia, Germany, Iran, Malaysia, Nigeria, Pakistan, Poland, Saudi Arabia, South Africa, Ukraine, and Venezuela) whereas 9 countries stayed sublinear (Brazil, Egypt, France, India, Japan, Mexico, Russia, Taiwan, and Uzbekistan). In contrast, Italy changed from superlinear to linear, and 12 countries changed from linear to sublinear (Argentina, Australia, Bangladesh, Canada, China, Indonesia, South Korea, Spain, Turkey, United States, United Kingdom).

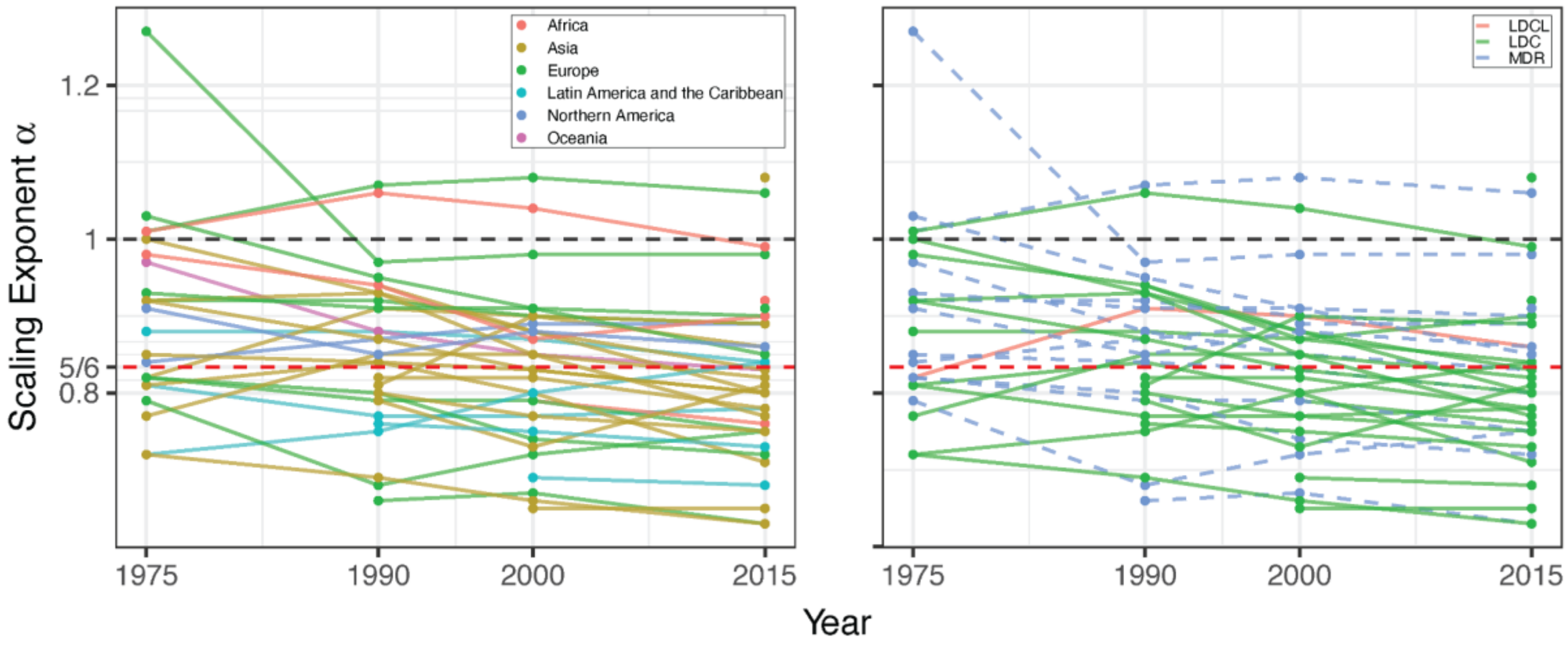

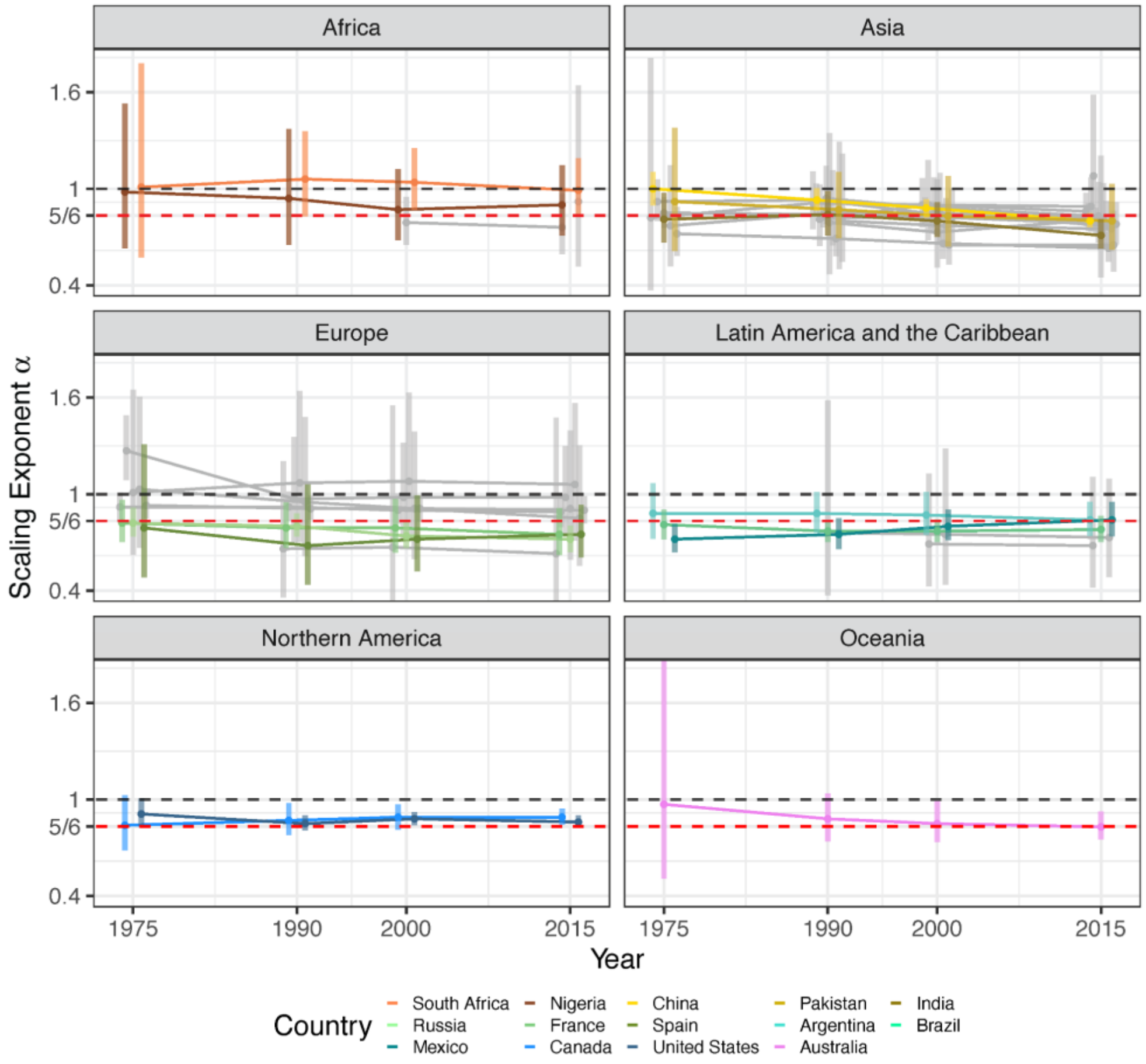

**Figure 3. Scaling exponents over time by: A) all regions, B) 2015 economic development status, C) Africa, D) Asia, E) Europe, F) Latin America and the Caribbean, G) Northern America, H) Oceania. Note each line represents the exponent for a country over time and vertical lines 95% Confidence Intervals. Select countries listed in the legend are highlighted.**

## Discussion

Using a new multi-decadal global-scale dataset, our results fill a gap in the literature by evaluating how commonly density increases with urban population size across the globe, leading to sublinear scaling of area with population size. Our results show mixed support for the UST consistent with previous studies (Batty and Furgeson 2011; Ortman et al. 2020). Sublinear scaling of area with population (increasing density) with an exponent of ⅚ is generalizable for only ~47% of countries globally. In contrast, ~48% of countries showed scaling relations statistically indistinguishable from constant densities across cities. Some countries (N =13) exhibited constant densities across cities for four time periods spanning three decades. Thus, we conclude that sublinear scaling as predicted by the UST is not the norm.

**Table 1. The number and percentage of countries by 95% Confidence Intervals of scaling exponents for 2015 in relation to Urban Scaling Theory predicted values (α = 5/6) and isometry (α = 1).**

| | RMA Total | OLS Total |
|---|---|---|
| # of countries containing both 5/6 and 1 | 17 (0.45%) | 12 (0.32%) |
| # of countries only containing 5/6 | 18 (0.47%) | 21 (0.55%) |
| # of countries that have neither | 3 (0.08%) | 5 (0.13%) |

Why do we see variation in scaling parameters and a lack of support for UST in many countries? It is unlikely a statistical artifact since several of the countries not supporting UST predictions have sufficient (>10) cities. UST assumes social-mixing and incremental development (Ortman et al. 2020). It is possible that the countries not scaling to the ⅚ UST prediction fail to meet these assumptions. In UST, the proposed mechanisms underlying increasingly denser larger cities are based on the capacity of the transport and communication system to connect all of its inhabitants and are assumed self-similar, maintaining their properties regardless of scale (Bettencourt 2013; 2020; Ortman and Lobo, 2020). However, such communication systems are costly, and the "ideal" (theoretical) connection proposed by UST may not be achieved in some countries. Countries with low per capita GDP may face greater challenges in investing and maintaining the needed transportation and communication connectivity costs for their inhabitants, potentially leading to lower-performing cities economically (Ortman et al. 2020). Data for physical infrastructure and economic and social variables are available for most high-income countries. More and better quality data across cities, especially from the "Global South," will allow us to better understand the links between city scalings, economic development, and country-level outcomes. Our results showing variation in urban density scaling for 38 countries is a step in that direction.

Only recently has the (normalization coefficient) of area-size scaling relations ($Y_0$) begun receiving attention in the UST literature (e.g., Bettencourt, 2020, Bettencourt et al., 2020, Ortman and Lobo, 2020). Our results show that $Y_0$ varies ~1 order of magnitude between countries and was related to socioeconomic and geographic drivers. Our results show greater variation in densities between countries and regions than within, with more developed economies having lower urban densities and occurring at higher latitudes. Interestingly, this latitudinal variation in urban densities and economic wealth mirrors gradients in biodiversity, with tropical areas having higher numbers of species of animals and plants than temperate regions having consequences for biodiversity. More investigation into the country- and region-specific drivers and consequences of variation in urban density scaling are thus clearly needed. Our results suggest that UST may need to be further refined and expanded to address this variation in scaling exponents and elevations theoretically.

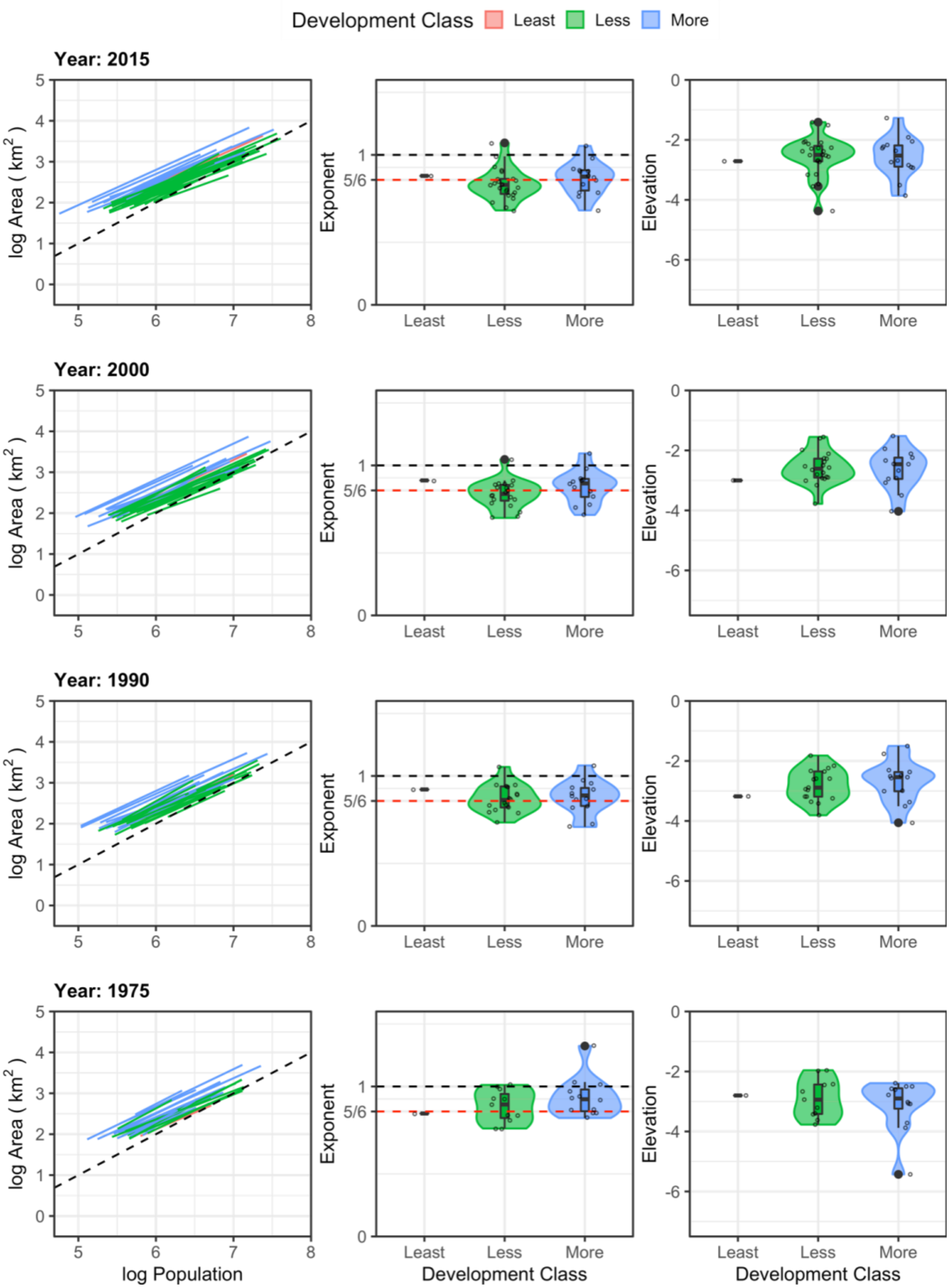

**Figure 4. Comparison of log area~ log population scaling for GHS dataset for different decades. Black dashed lines represent the 1/1 line (e.g., $\alpha$ = 1). All regressions (Reduced Major Axis) are country-level with at least 5 cities of > 50,000 people and >= 100 km$^2$ for the specific year.**

*Urban density and biodiversity*

Understanding the scaling relations of urban densities has important consequences for green space and habitat for biodiversity (Uchida et al. 2021; Dunn et al. 2022). The country-specific parameter estimates in scaling exponents and constants (elevation) provided in the Appendix have implications for country-specific management and policy. For instance, population density and city area are both fundamental variables for biodiversity conservation, influencing human pressures on habitats and the fragmented structure of habitats. If density increases with city size for a given country, different incentives may be put in place to allow compensatory green space in larger, denser cities. The scaling of green space, therefore, may impact biodiversity.

Uchida et al. (2021) hypothesized that urban areas have higher biodiversity than similar size non-urban areas where the scaling exponent of the Species-Area-Relationship may be steeper with higher log-log intercepts. However, this relationship has largely been investigated in cities of the temperate Northern Hemisphere. For example, in European cities, bird species richness appears to increases with city area as power-law with exponent 0.18. Given the shallow exponent, however, only a small amount of species turnover is required in order for small cities to support more biodiversity per inhabitant. Indeed, for specific taxa like birds, species-area richness is found to be higher in urban areas compared to their rural counterparts (Ferenc et al. 2016; Callaghan et al. 2018; but see Fattorini et al. 2018 in Italy) and also depends on latitude (Murthy et al. 2016). However, this outcome may not necessarily be congruent with conservation priorities since urban areas often have greater numbers of species of both native and non-native generalist species resulting in part due increased species introductions and resource availability by humans in urban environments (Murthy et al. 2016; Ferenc et al. 2016). Our study highlights a latitudinal gradient in population density where the highest density cities in the world are located in the tropics, where there is generally higher terrestrial biodiversity (Brown, 2014). Further studies are needed to understand how variation in density scalings impacts the species-area relationship in cities in temperate versus tropics cities and in different biomes (Uchida et al. 2021).

More broadly, our results highlight the need to better understand how other large-scale geographic, political, and socioeconomic factors drive variation in population density and shape the area-size scaling exponent ($\alpha$) and elevation ( $Y_0$). In countries in which the scaling of area with population size is linear, the effect of the size distribution of cities should be invariant for regional biodiversity. Yet, other attributes of cities for urban biodiversity could also be important that do not scale linearly with city size. Additionally, whether or not multiple small cities can support more total urban species richness (gamma diversity) than a few large cities of the equivalent population depends on the rate of turnover in species composition between cities and the exponent of the species-area relationship of cities (Uchida et al., 2021). Thus, resolving this dilemma requires a better understanding of the scaling of various biodiversity-relevant features of cities, such as how green space, biodiversity, net primary productivity, and habitat connectivity scale with city size. Additionally, the spatial distribution of cities of varying sizes undoubtedly affects the regional connectivity of habitats, influencing rates of colonization and dispersal between and within cities and rural areas. Our analysis of city area and population provide crucial information to advance more empirical and theoretical development to effectively understanding country-level scaling effects.

*Extra-urban scalings*

Cities are dependent on extra-urban interactions from regional and global trade networks (Burger et al. 2012) not currently considered in UST. The productivity of their boundaries does not constrain modern cities due to the industrial revolution, and the use of fossil fuels to now import vast quantities of goods and services from global trade networks into cities (Krausmann et al., 2008; Meyfroidt et al., 2011, Burger et al., 2012, 2019; Schramski et al., 2015, 2019, 2020). An increase in urbanization within countries over time shows (i) greater dependence on the net import of raw materials from abroad, increasing the *actual* footprint of the city that is not being taken into account (Rees and Wackernagl 1996; Hidalgo, 2009; Meyfroidt et al., 2011; Burger et al., 2012; Schramski et al., 2019, 2020), (ii) a general trend of per capita increase in extra-metabolic and metabolic (i.e., food) energy sources and material consumption (Brown et al., 2011; Burger et al., 2017, 2019) (iii) an accelerated extinction of biodiversity and alterations of ecosystem function (e.g., Ceballos et al., 2015) that are so tightly related to the uncertainties behind the global change and the future Anthropocene era scenarios and our survival as species (Barnosky et al., 2012; Steffen et al., 2018). Future studies should advance linking urban scaling relations to extra-urban dynamics that are currently not accounted for in theoretical and empirical studies of the UST.

**Conclusion**

Using a global database of urban area and population size, we show the variation in population densities and scaling relationships between urban area and population size across the spectrum of socioeconomic development on Earth. Support for UST expectations of ⅚ scaling exponents was mixed, with nearly half of countries exhibiting scaling indistinguishable from linear and a surprising substantial number of countries maintaining this apparent linearity and invariance of population density over three decades. Further research is thus needed to understand the interactions among economics, demography, history, and culture that may underlie these context-specific relationships. While our study involves a wealth of data, including hundreds of cities and scaling analysis up to 38 countries over 4 time periods, we have only focused on area and population. Additional variables must be investigated at these large cross-country and cross-city scales. Whether or not a single general theory can manage urban biodiversity given the variation in urban density scaling uncovered here remains an open question.

**Funding**

This project was supported by funding from the Bridging Biodiversity and Conservation Science Program of the Arizona Institutes of Resilience at the University of Arizona, Postdoctoral funding from Northeastern University, a Lyman T Johnson Diversity Fellowship from the University of Kentucky's Graduate School, and a Postdoctoral Fellowship from Agencia Nacional de Investigación y Desarrollo de Chile ANID 3190609.

**Data Availability Statement**

The original contributions presented in the study are included in the article/Supplementary Material, further inquiries can be directed to the corresponding author/s.

**Author Contributions**

JRB led the conceptualization, execution, analysis, discussion, and writing with input from all authors. JRB, JGO, IH, VPW, MS, KJL, TB, ARC, XF, CHH, ASMGK contributed to data analysis, interpretation of results and discussion. All authors contributed to editing the article and approved the submitted version.

**Appendix 1: Summary OLS and RMA statistics of scalings of log area with log population for countries >5 cities of >100 km² and >50,000 population, per time frame (N = 933 cities in 2015, 738 in 2000, 591 in 1990, 416 in 1975).**

| Region | Country (n = cities) | Dev. Class | Year | OLS | | RMA | | P-value | $R^2$ |
|---|---|---|---|---|---|---|---|---|---|
| | | | | Exponent [95% CI] | Elevation [95% CI] | Exponent [95% CI] | Elevation [95% CI] | | |
| Sublinear (RMA Exp. CI < 1) | | | | | | | | | |
| Europe | Russia (44) | MDR | 2015 | 0.67 [0.58, 0.75] | -3.72 [-4.83, -2.62] | 0.72 [0.64, 0.81] | -4.40 [-5.50, -3.30] | 0.000 | 0.86 |
| Asia | Japan (33) | MDR | 2015 | 0.77 [0.69, 0.85] | -4.73 [-5.79, -3.66] | 0.80 [0.72, 0.88] | -5.12 [-6.19, -4.06] | 0.000 | 0.93 |
| Northern America | United States (151) | MDR | 2015 | 0.83 [0.80, 0.87] | -5.04 [-5.51, -4.56] | 0.86 [0.83, 0.90] | -5.43 [-5.90, -4.95] | 0.000 | 0.93 |
| Europe | France (11) | MDR | 2015 | 0.73 [0.58, 0.87] | -4.37 [-6.36, -2.37] | 0.75 [0.62, 0.91] | -4.71 [-6.71, -2.72] | 0.000 | 0.93 |
| Oceania | Australia (8) | MDR | 2015 | 0.83 [0.74, 0.92] | -5.18 [-6.36, -4.01] | 0.83 [0.75, 0.92] | -5.24 [-6.42, -4.07] | 0.000 | 0.99 |
| Europe | Spain (11) | MDR | 2015 | 0.72 [0.56, 0.88] | -4.56 [-6.80, -2.32] | 0.75 [0.61, 0.93] | -4.99 [-7.23, -2.75] | 0.000 | 0.92 |
| Northern America | Canada (14) | MDR | 2015 | 0.89 [0.84, 0.94] | -6.26 [-6.92, -5.60] | 0.89 [0.85, 0.94] | -6.30 [-6.96, -5.64] | 0.000 | 0.99 |
| Europe | United Kingdom (23) | MDR | 2015 | 0.88 [0.79, 0.96] | -6.45 [-7.58, -5.32] | 0.90 [0.82, 0.99] | -6.71 [-7.84, -5.58] | 0.000 | 0.96 |
| Asia | Taiwan (7) | LDC | 2015 | 0.63 [0.54, 0.72] | -3.18 [-4.47, -1.90] | 0.63 [0.55, 0.73] | -3.25 [-4.54, -1.97] | 0.000 | 0.98 |
| Asia | India (100) | LDC | 2015 | 0.55 [0.46, 0.64] | -2.43 [-3.73, -1.12] | 0.71 [0.63, 0.81] | -4.78 [-6.08, -3.48] | 0.000 | 0.59 |
| Asia | China (172) | LDC | 2015 | 0.75 [0.70, 0.79] | -5.02 [-5.64, -4.40] | 0.80 [0.76, 0.85] | -5.79 [-6.41, -5.18] | 0.000 | 0.87 |
| Asia | Uzbekistan (6) | LDC | 2015 | 0.63 [0.45, 0.82] | -3.28 [-5.83, -0.72] | 0.65 [0.49, 0.86] | -3.47 [-6.02, -0.92] | 0.001 | 0.96 |
| Latin America and | Brazil (36) | LDC | 2015 | 0.74 [0.66, | -4.97 [-6.12, | 0.78 [0.70, | -5.48 [-6.63, | 0.00 | 0.91 |
| the Caribbean | | | | 0.83] | -3.82] | 0.87] | -4.33] | 0 | |
| Asia | Indonesia (50) | LDC | 2015 | 0.77 [0.68, 0.85] | -5.04 [-6.21, -3.87] | 0.82 [0.74, 0.91] | -5.78 [-6.95, -4.61] | 0.000 | 0.87 |
| Asia | Vietnam (10) | LDC | 2015 | 0.76 [0.63, 0.89] | -4.98 [-6.79, -3.17] | 0.78 [0.66, 0.92] | -5.21 [-7.02, -3.40] | 0.000 | 0.96 |
| Asia | Turkey (13) | LDC | 2015 | 0.71 [0.56, 0.87] | -4.70 [-6.88, -2.51] | 0.75 [0.61, 0.92] | -5.22 [-7.41, -3.04] | 0.000 | 0.9 |
| Asia | South Korea (9) | LDC | 2015 | 0.76 [0.62, 0.90] | -5.26 [-7.21, -3.30] | 0.77 [0.65, 0.92] | -5.47 [-7.43, -3.52] | 0.000 | 0.96 |
| Latin America and the Caribbean | Mexico (28) | LDC | 2015 | 0.80 [0.69, 0.90] | -5.71 [-7.17, -4.24] | 0.84 [0.74, 0.95] | -6.29 [-7.75, -4.82] | 0.000 | 0.9 |
| Latin America and the Caribbean | Argentina (7) | LDC | 2015 | 0.83 [0.73, 0.94] | -6.14 [-7.63, -4.65] | 0.84 [0.74, 0.95] | -6.21 [-7.70, -4.72] | 0.000 | 0.99 |
| Africa | Egypt (13) | LDC | 2015 | 0.70 [0.52, 0.89] | -4.54 [-7.13, -1.95] | 0.76 [0.59, 0.97] | -5.28 [-7.87, -2.69] | 0.000 | 0.86 |
| Asia | Bangladesh (33) | LDCL | 2015 | 0.82 [0.71, 0.92] | -5.59 [-7.00, -4.18] | 0.86 [0.77, 0.97] | -6.24 [-7.64, -4.83] | 0.000 | 0.89 |
| Linear (RMA Exp. CI = 1) | | | | | | | | | |
| Europe | Germany (20) | MDR | 2015 | 0.87 [0.76, 0.97] | -6.20 [-7.59, -4.81] | 0.89 [0.79, 1.00] | -6.53 [-7.92, -5.14] | 0.000 | 0.95 |
| Europe | Italy (8) | MDR | 2015 | 0.94 [0.66, 1.22] | -7.50 [-11.40, -3.61] | 0.98 [0.74, 1.30] | -8.07 [-11.97, -4.18] | 0.000 | 0.92 |
| Europe | Ukraine (11) | MDR | 2015 | 0.69 [0.32, 1.06] | -4.02 [-9.11, 1.06] | 0.85 [0.56, 1.30] | -6.19 [-11.28, -1.10] | 0.002 | 0.66 |
| Europe | Netherlands (5) | MDR | 2015 | 0.88 [0.48, 1.28] | -6.39 [-11.72, -1.06] | 0.91 [0.59, 1.39] | -6.74 [-12.08, -1.41] | 0.006 | 0.94 |
| Europe | Belgium (5) | MDR | 2015 | 0.53 [-0.07, 1.14] | -1.70 [-9.64, 6.25] | 0.63 [0.27, 1.48] | -2.92 [-10.87, 5.02] | 0.068 | 0.72 |
| Europe | Poland (7) | MDR | 2015 | 0.99 [0.56, 1.42] | -7.97 [-13.77, -2.18] | 1.06 [0.71, 1.57] | -8.89 [-14.68, -3.09] | 0.002 | 0.88 |
| Asia | Pakistan (13) | LDC | 2015 | 0.74 [0.53, 0.94] | -5.26 [-8.27, -2.24] | 0.80 [0.62, 1.03] | -6.17 [-9.19, -3.16] | 0.000 | 0.85 |
| Asia | Saudi Arabia (10) | LDC | 2015 | 0.77 [0.56, 0.97] | -5.21 [-8.04, -2.38] | 0.81 [0.63, 1.04] | -5.76 [-8.59, -2.93] | 0.000 | 0.9 |
| Asia | Malaysia (9) | LDC | 2015 | 0.81 [0.62, | -5.52 [-8.05, | 0.83 [0.67, | -5.89 [-8.41, | 0.00 | 0.94 |
| | | | | 0.99] | -3.00] | 1.04] | -3.36] | 0 | |
| Asia | Iran (11) | LDC | 2015 | 0.86 [0.69, 1.03] | -6.85 [-9.22, -4.47] | 0.89 [0.73, 1.07] | -7.25 [-9.62, -4.87] | 0.000 | 0.94 |
| Latin America and the Caribbean | Venezuela (7) | LDC | 2015 | 0.68 [0.37, 0.98] | -4.23 [-8.51, 0.04] | 0.73 [0.48, 1.09] | -4.94 [-9.21, -0.66] | 0.002 | 0.87 |
| Latin America and the Caribbean | Colombia (5) | LDC | 2015 | 0.66 [0.31, 1.00] | -4.33 [-9.40, 0.75] | 0.68 [0.42, 1.11] | -4.71 [-9.78, 0.36] | 0.009 | 0.93 |
| Africa | Nigeria (19) | LDC | 2015 | 0.80 [0.58, 1.01] | -5.75 [-8.84, -2.67] | 0.90 [0.71, 1.15] | -7.26 [-10.35, -4.18] | 0.000 | 0.78 |
| Africa | South Africa (8) | LDC | 2015 | 0.97 [0.79, 1.16] | -7.93 [-10.50, -5.36] | 0.99 [0.83, 1.19] | -8.17 [-10.74, -5.60] | 0.000 | 0.97 |
| Asia | Philippines (5) | LDC | 2015 | 0.71 [0.33, 1.08] | -4.39 [-9.88, 1.10] | 0.73 [0.45, 1.20] | -4.81 [-10.30, 0.68] | 0.010 | 0.92 |
| Asia | Iraq (6) | LDC | 2015 | 1.04 [0.61, 1.46] | -9.44 [-15.43, -3.46] | 1.08 [0.74, 1.58] | -10.07 [-16.05, -4.08] | 0.002 | 0.92 |
| Africa | Morocco (5) | LDC | 2015 | 0.87 [0.31, 1.43] | -7.24 [-15.26, 0.79] | 0.92 [0.52, 1.64] | -7.98 [-16.00, 0.04] | 0.016 | 0.89 |
| Sublinear (RMA Exp. CI < 1) | | | | | | | | | |
| Europe | Russia (42) | MDR | 2000 | 0.70 [0.61, 0.78] | -4.22 [-5.34, -3.11] | 0.74 [0.67, 0.83] | -4.85 [-5.97, -3.74] | 0.000 | 0.88 |
| Asia | Japan (30) | MDR | 2000 | 0.80 [0.71, 0.89] | -5.22 [-6.40, -4.04] | 0.83 [0.75, 0.92] | -5.64 [-6.82, -4.46] | 0.000 | 0.93 |
| Northern America | United States (123) | MDR | 2000 | 0.85 [0.81, 0.89] | -5.27 [-5.81, -4.74] | 0.88 [0.84, 0.92] | -5.67 [-6.20, -5.13] | 0.000 | 0.93 |
| Europe | France (11) | MDR | 2000 | 0.75 [0.59, 0.92] | -4.73 [-6.99, -2.48] | 0.79 [0.64, 0.97] | -5.16 [-7.41, -2.90] | 0.000 | 0.92 |
| Northern America | Canada (11) | MDR | 2000 | 0.88 [0.80, 0.96] | -6.06 [-7.13, -4.99] | 0.89 [0.81, 0.97] | -6.15 [-7.22, -5.08] | 0.000 | 0.99 |
| Oceania | Australia (8) | MDR | 2000 | 0.84 [0.72, 0.97] | -5.26 [-6.96, -3.56] | 0.85 [0.73, 0.99] | -5.39 [-7.09, -3.69] | 0.000 | 0.98 |
| Europe | Spain (8) | MDR | 2000 | 0.68 [0.44, 0.91] | -3.94 [-7.20, -0.69] | 0.72 [0.52, 0.99] | -4.49 [-7.75, -1.24] | 0.000 | 0.89 |
| Europe | United Kingdom (21) | MDR | 2000 | 0.89 [0.81, 0.97] | -6.58 [-7.67, -5.48] | 0.91 [0.83, 0.99] | -6.79 [-7.89, -5.70] | 0.000 | 0.96 |
| Asia | Taiwan (7) | LDC | 2000 | 0.65 [0.55, 0.75] | -3.57 [-4.96, -2.18] | 0.66 [0.57, 0.76] | -3.65 [-5.04, -2.26] | 0.000 | 0.98 |
| Asia | Uzbekistan (6) | LDC | 2000 | 0.65 [0.51, 0.79] | -3.46 [-5.32, -1.60] | 0.65 [0.53, 0.81] | -3.56 [-5.42, -1.70] | 0.000 | 0.98 |
| Latin America and the Caribbean | Brazil (36) | LDC | 2000 | 0.74 [0.67, 0.81] | -4.88 [-5.86, -3.91] | 0.77 [0.70, 0.84] | -5.26 [-6.23, -4.28] | 0.000 | 0.93 |
| Latin America and the Caribbean | Mexico (25) | LDC | 2000 | 0.77 [0.68, 0.87] | -5.34 [-6.65, -4.03] | 0.80 [0.71, 0.90] | -5.77 [-7.07, -4.46] | 0.000 | 0.92 |
| Asia | India (71) | LDC | 2000 | 0.66 [0.56, 0.77] | -4.13 [-5.67, -2.59] | 0.80 [0.70, 0.92] | -6.10 [-7.64, -4.56] | 0.000 | 0.69 |
| Asia | China (129) | LDC | 2000 | 0.82 [0.76, 0.87] | -6.07 [-6.87, -5.28] | 0.88 [0.82, 0.94] | -6.93 [-7.72, -6.14] | 0.000 | 0.87 |
| Asia | Turkey (10) | LDC | 2000 | 0.75 [0.60, 0.90] | -5.16 [-7.30, -3.02] | 0.77 [0.63, 0.94] | -5.48 [-7.62, -3.34] | 0.000 | 0.94 |
| Asia | Indonesia (36) | LDC | 2000 | 0.82 [0.73, 0.90] | -5.80 [-7.00, -4.59] | 0.85 [0.77, 0.94] | -6.31 [-7.51, -5.10] | 0.000 | 0.92 |
| Africa | Egypt (7) | LDC | 2000 | 0.78 [0.63, 0.92] | -5.68 [-7.80, -3.55] | 0.79 [0.65, 0.95] | -5.83 [-7.95, -3.71] | 0.000 | 0.97 |
| Asia | Vietnam (10) | LDC | 2000 | 0.80 [0.63, 0.96] | -5.58 [-7.79, -3.37] | 0.82 [0.67, 1.00] | -5.91 [-8.12, -3.70] | 0.000 | 0.94 |
| Linear (RMA Exp. CI = 1) | | | | | | | | | |
| Europe | Germany (19) | MDR | 2000 | 0.89 [0.79, 0.98] | -6.48 [-7.74, -5.22] | 0.90 [0.82, 1.00] | -6.73 [-7.99, -5.47] | 0.000 | 0.96 |
| Europe | Italy (8) | MDR | 2000 | 0.93 [0.63, 1.23] | -7.41 [-11.55, -3.27] | 0.98 [0.72, 1.32] | -8.05 [-12.20, -3.91] | 0.000 | 0.91 |
| Europe | Ukraine (11) | MDR | 2000 | 0.74 [0.35, 1.14] | -4.81 [-10.23, 0.61] | 0.91 [0.60, 1.39] | -7.09 [-12.51, -1.67] | 0.002 | 0.67 |
| Europe | Belgium (5) | MDR | 2000 | 0.57 [-0.06, 1.21] | -2.25 [-10.49, 5.98] | 0.67 [0.29, 1.55] | -3.49 [-11.73, 4.74] | 0.063 | 0.74 |
| Europe | Poland (7) | MDR | 2000 | 1.00 [0.54, 1.46] | -8.23 [-14.52, -1.94] | 1.08 [0.71, 1.63] | -9.29 [-15.58, -2.99] | 0.003 | 0.86 |
| Latin America and the Caribbean | Argentina (7) | LDC | 2000 | 0.86 [0.73, 1.00] | -6.53 [-8.41, -4.65] | 0.87 [0.75, 1.01] | -6.64 [-8.52, -4.76] | 0.000 | 0.98 |
| Asia | Saudi Arabia (7) | LDC | 2000 | 0.69 [0.42, 0.97] | -4.02 [-7.82, -0.21] | 0.73 [0.51, 1.06] | -4.57 [-8.38, -0.76] | 0.001 | 0.89 |
| Asia | Malaysia (8) | LDC | 2000 | 0.86 [0.68, 1.04] | -6.22 [-8.54, -3.90] | 0.88 [0.72, 1.07] | -6.46 [-8.78, -4.14] | 0.000 | 0.96 |
| Asia | Pakistan (8) | LDC | 2000 | 0.80 [0.58, 1.02] | -6.23 [-9.50, -2.96] | 0.83 [0.64, 1.08] | -6.67 [-9.93, -3.40] | 0.000 | 0.93 |
| Asia | South Korea (7) | LDC | 2000 | 0.83 [0.60, 1.05] | -6.40 [-9.65, -3.15] | 0.85 [0.65, 1.10] | -6.73 [-9.98, -3.48] | 0.000 | 0.95 |
| Africa | Nigeria (16) | LDC | 2000 | 0.78 [0.56, 1.00] | -5.55 [-8.61, -2.48] | 0.87 [0.68, 1.12] | -6.79 [-9.86, -3.73] | 0.000 | 0.81 |
| Latin America and the Caribbean | Colombia (5) | LDC | 2000 | 0.67 [0.31, 1.02] | -4.47 [-9.60, 0.67] | 0.69 [0.43, 1.13] | -4.86 [-9.99, 0.27] | 0.009 | 0.92 |
| Asia | Iran (7) | LDC | 2000 | 0.87 [0.62, 1.12] | -6.89 [-10.45, -3.32] | 0.90 [0.68, 1.18] | -7.27 [-10.84, -3.70] | 0.000 | 0.94 |
| Africa | South Africa (8) | LDC | 2000 | 1.03 [0.83, 1.22] | -8.46 [-11.12, -5.80] | 1.04 [0.87, 1.25] | -8.71 [-11.36, -6.05] | 0.000 | 0.97 |
| Latin America and the Caribbean | Venezuela (6) | LDC | 2000 | 0.68 [0.26, 1.11] | -4.29 [-10.22, 1.64] | 0.75 [0.44, 1.29] | -5.20 [-11.13, 0.73] | 0.011 | 0.83 |
| Asia | Bangladesh (18) | LDCL | 2000 | 0.85 [0.70, 1.01] | -6.23 [-8.37, -4.08] | 0.90 [0.76, 1.07] | -6.90 [-9.04, -4.75] | 0.000 | 0.9 |
| Sublinear (RMA Exp. CI < 1) | | | | | | | | | |
| Europe | Russia (38) | MDR | 1990 | 0.78 [0.71, 0.85] | -5.39 [-6.35, -4.43] | 0.80 [0.74, 0.88] | -5.77 [-6.73, -4.81] | 0.000 | 0.93 |
| Northern America | United States (97) | MDR | 1990 | 0.82 [0.78, 0.87] | -5.06 [-5.66, -4.46] | 0.85 [0.81, 0.90] | -5.45 [-6.05, -4.85] | 0.000 | 0.93 |
| Asia | Japan (30) | MDR | 1990 | 0.81 [0.72, 0.90] | -5.43 [-6.63, -4.23] | 0.84 [0.76, 0.93] | -5.85 [-7.05, -4.65] | 0.000 | 0.93 |
| Europe | France (11) | MDR | 1990 | 0.77 [0.63, 0.91] | -5.02 [-6.94, -3.10] | 0.79 [0.66, 0.95] | -5.32 [-7.24, -3.41] | 0.000 | 0.94 |
| Northern America | Canada (11) | MDR | 1990 | 0.86 [0.76, 0.96] | -5.78 [-7.11, -4.45] | 0.87 [0.78, 0.98] | -5.91 [-7.24, -4.59] | 0.000 | 0.98 |
| Europe | United Kingdom (21) | MDR | 1990 | 0.90 [0.82, 0.98] | -6.69 [-7.71, -5.68] | 0.91 [0.84, 0.99] | -6.88 [-7.89, -5.86] | 0.000 | 0.97 |
| Asia | Taiwan (7) | LDC | 1990 | 0.69 [0.56, 0.81] | -4.08 [-5.79, -2.37] | 0.69 [0.58, 0.83] | -4.19 [-5.90, -2.48] | 0.000 | 0.98 |
| Latin America and the Caribbean | Brazil (30) | LDC | 1990 | 0.75 [0.68, 0.81] | -4.90 [-5.82, -3.99] | 0.77 [0.70, 0.84] | -5.17 [-6.08, -4.25] | 0.000 | 0.95 |
| Latin America and the Caribbean | Mexico (22) | LDC | 1990 | 0.72 [0.63, 0.82] | -4.58 [-5.89, -3.28] | 0.75 [0.66, 0.85] | -4.97 [-6.28, -3.67] | 0.000 | 0.93 |
| Asia | India (51) | LDC | 1990 | 0.68 [0.55, 0.82] | -4.48 [-6.46, -2.51] | 0.84 [0.71, 0.98] | -6.65 [-8.63, -4.67] | 0.000 | 0.67 |
| Asia | China (99) | LDC | 1990 | 0.87 [0.80, 0.93] | -6.81 [-7.79, -5.83] | 0.93 [0.86, 1.00] | -7.74 [-8.71, -6.76] | 0.000 | 0.86 |
| Linear (RMA Exp. CI = 1) | | | | | | | | | |
| Europe | Germany (19) | MDR | 1990 | 0.90 [0.81, 0.99] | -6.70 [-7.94, -5.46] | 0.92 [0.83, 1.02] | -6.93 [-8.17, -5.69] | 0.000 | 0.96 |
| Oceania | Australia (8) | MDR | 1990 | 0.86 [0.71, 1.01] | -5.54 [-7.52, -3.55] | 0.88 [0.74, 1.04] | -5.71 [-7.69, -3.72] | 0.000 | 0.97 |
| Europe | Spain (7) | MDR | 1990 | 0.62 [0.31, 0.94] | -3.27 [-7.60, 1.06] | 0.68 [0.44, 1.06] | -4.05 [-8.38, 0.28] | 0.004 | 0.84 |
| Europe | Belgium (5) | MDR | 1990 | 0.61 [0.19, 1.04] | -2.90 [-8.41, 2.61] | 0.66 [0.36, 1.21] | -3.45 [-8.96, 2.06] | 0.019 | 0.88 |
| Europe | Italy (8) | MDR | 1990 | 0.92 [0.59, 1.24] | -7.29 [-11.85, -2.73] | 0.97 [0.70, 1.35] | -8.08 [-12.64, -3.52] | 0.000 | 0.89 |
| Europe | Ukraine (11) | MDR | 1990 | 0.76 [0.33, 1.20] | -5.15 [-11.08, 0.78] | 0.95 [0.62, 1.48] | -7.77 [-13.70, -1.84] | 0.003 | 0.64 |
| Europe | Poland (7) | MDR | 1990 | 0.99 [0.52, 1.46] | -8.25 [-14.68, -1.83] | 1.07 [0.70, 1.64] | -9.35 [-15.78, -2.93] | 0.003 | 0.86 |
| Latin America and the Caribbean | Argentina (7) | LDC | 1990 | 0.87 [0.74, 1.00] | -6.67 [-8.44, -4.91] | 0.88 [0.76, 1.01] | -6.77 [-8.54, -5.00] | 0.000 | 0.98 |
| Asia | Iran (6) | LDC | 1990 | 0.80 [0.61, 0.99] | -5.80 [-8.48, -3.13] | 0.81 [0.64, 1.03] | -5.97 [-8.64, -3.29] | 0.000 | 0.97 |
| Asia | Indonesia (29) | LDC | 1990 | 0.90 [0.80, 0.99] | -7.00 [-8.38, -5.62] | 0.93 [0.84, 1.03] | -7.47 [-8.85, -6.10] | 0.000 | 0.93 |
| Asia | Pakistan (7) | LDC | 1990 | 0.86 [0.65, 1.06] | -7.06 [-10.09, -4.03] | 0.87 [0.69, 1.10] | -7.33 [-10.37, -4.30] | 0.000 | 0.96 |
| Asia | South Korea (7) | LDC | 1990 | 0.83 [0.59, 1.06] | -6.48 [-9.87, -3.08] | 0.85 [0.65, 1.11] | -6.84 [-10.23, -3.44] | 0.000 | 0.94 |
| Asia | Malaysia (5) | LDC | 1990 | 0.94 [0.76, 1.12] | -7.27 [-9.67, -4.86] | 0.94 [0.78, 1.14] | -7.34 [-9.74, -4.93] | 0.000 | 0.99 |
| Asia | Vietnam (5) | LDC | 1990 | 0.80 [0.46, 1.13] | -5.72 [-10.41, -1.02] | 0.82 [0.55, 1.22] | -6.00 [-10.70, -1.31] | 0.005 | 0.95 |
| Asia | Turkey (5) | LDC | 1990 | 0.77 [0.38, 1.17] | -5.55 [-11.21, 0.10] | 0.80 [0.50, 1.29] | -5.97 [-11.63, -0.32] | 0.008 | 0.93 |
| Asia | Saudi Arabia (7) | LDC | 1990 | 0.70 [0.26, 1.13] | -4.05 [-10.02, 1.92] | 0.79 [0.47, 1.34] | -5.38 [-11.34, 0.59] | 0.009 | 0.77 |
| Africa | South Africa (7) | LDC | 1990 | 1.03 [0.77, 1.30] | -8.42 [-12.05, -4.78] | 1.06 [0.83, 1.36] | -8.76 [-12.40, -5.13] | 0.000 | 0.95 |
| Africa | Nigeria (8) | LDC | 1990 | 0.87 [0.51, 1.23] | -6.85 [-11.97, -1.74] | 0.94 [0.65, 1.37] | -7.87 [-12.98, -2.75] | 0.001 | 0.85 |
| Latin America and the Caribbean | Venezuela (5) | LDC | 1990 | 0.69 [0.08, 1.30] | -4.36 [-12.82, 4.11] | 0.76 [0.37, 1.58] | -5.40 [-13.87, 3.06] | 0.036 | 0.81 |
| Asia | Bangladesh (11) | LDCL | 1990 | 0.88 [0.70, 1.06] | -6.86 [-9.42, -4.29] | 0.91 [0.75, 1.11] | -7.31 [-9.88, -4.75] | 0.000 | 0.93 |
| Sublinear (RMA Exp. CI < 1) | | | | | | | | | |
| Europe | Russia (30) | MDR | 1975 | 0.79 [0.70, 0.88] | -5.55 [-6.77, -4.33] | 0.82 [0.74, 0.92] | -6.00 [-7.22, -4.79] | 0.000 | 0.92 |
| Asia | Japan (29) | MDR | 1975 | 0.81 [0.72, 0.91] | -5.45 [-6.75, -4.15] | 0.85 [0.76, 0.95] | -5.93 [-7.23, -4.63] | 0.000 | 0.92 |
| Europe | France (11) | MDR | 1975 | 0.80 [0.67, 0.94] | -5.52 [-7.27, -3.76] | 0.82 [0.70, 0.96] | -5.76 [-7.52, -4.01] | 0.000 | 0.96 |
| Latin America and the Caribbean | Mexico (15) | LDC | 1975 | 0.71 [0.62, 0.79] | -4.35 [-5.54, -3.15] | 0.72 [0.64, 0.81] | -4.55 [-5.74, -3.36] | 0.000 | 0.96 |
| Asia | Taiwan (5) | LDC | 1975 | 0.71 [0.56, 0.87] | -4.44 [-6.58, -2.30] | 0.72 [0.58, 0.89] | -4.51 [-6.65, -2.36] | 0.001 | 0.99 |
| Latin America and the Caribbean | Brazil (20) | LDC | 1975 | 0.79 [0.69, 0.88] | -5.35 [-6.62, -4.08] | 0.81 [0.72, 0.90] | -5.65 [-6.92, -4.38] | 0.000 | 0.95 |
| Asia | India (35) | LDC | 1975 | 0.68 [0.52, 0.83] | -4.32 [-6.53, -2.11] | 0.81 [0.67, 0.97] | -6.16 [-8.37, -3.95] | 0.000 | 0.71 |
| Linear (RMA Exp. CI = 1) | | | | | | | | | |
| Northern America | United States (72) | MDR | 1975 | 0.84 [0.75, 0.92] | -5.38 [-6.54, -4.21] | 0.91 [0.83, 1.01] | -6.40 [-7.57, -5.23] | 0.000 | 0.84 |
| Europe | Germany (17) | MDR | 1975 | 0.90 [0.81, 0.99] | -6.76 [-7.96, -5.56] | 0.92 [0.83, 1.01] | -6.95 [-8.15, -5.75] | 0.000 | 0.97 |
| Europe | United Kingdom (20) | MDR | 1975 | 0.91 [0.83, 1.00] | -6.86 [-7.99, -5.73] | 0.93 [0.85, 1.02] | -7.07 [-8.20, -5.94] | 0.000 | 0.97 |
| Northern America | Canada (8) | MDR | 1975 | 0.82 [0.65, 0.99] | -5.25 [-7.57, -2.93] | 0.84 [0.68, 1.03] | -5.49 [-7.81, -3.17] | 0.000 | 0.96 |
| Europe | Spain (6) | MDR | 1975 | 0.74 [0.32, 1.15] | -4.94 [-10.74, 0.86] | 0.79 [0.48, 1.31] | -5.75 [-11.55, 0.05] | 0.008 | 0.86 |
| Europe | Ukraine (11) | MDR | 1975 | 0.83 [0.36, 1.30] | -6.10 [-12.49, 0.29] | 1.03 [0.67, 1.60] | -8.92 [-15.31, -2.54] | 0.003 | 0.64 |
| Europe | Poland (7) | MDR | 1975 | 0.91 [0.40, 1.42] | -7.15 [-14.15, -0.15] | 1.01 [0.62, 1.65] | -8.56 [-15.57, -1.56] | 0.006 | 0.81 |
| Oceania | Australia (5) | MDR | 1975 | 0.90 [0.23, 1.58] | -6.09 [-15.45, 3.26] | 0.97 [0.51, 1.86] | -7.09 [-16.44, 2.26] | 0.024 | 0.86 |
| Latin America and the Caribbean | Argentina (7) | LDC | 1975 | 0.86 [0.69, 1.04] | -6.58 [-8.97, -4.20] | 0.88 [0.72, 1.07] | -6.76 [-9.15, -4.38] | 0.000 | 0.97 |
| Asia | China (65) | LDC | 1975 | 0.91 [0.80, 1.01] | -7.42 [-8.89, -5.94] | 1.00 [0.90, 1.10] | -8.69 [-10.17, -7.22] | 0.000 | 0.83 |
| Asia | South Korea (5) | LDC | 1975 | 0.75 [0.44, 1.06] | -5.30 [-9.87, -0.74] | 0.77 [0.52, 1.15] | -5.58 [-10.14, -1.02] | 0.005 | 0.95 |
| Asia | Pakistan (6) | LDC | 1975 | 0.88 [0.49, 1.26] | -7.23 [-12.70, -1.77] | 0.92 [0.61, 1.38] | -7.84 [-13.30, -2.37] | 0.003 | 0.91 |
| Africa | Nigeria (6) | LDC | 1975 | 0.93 [0.48, 1.38] | -7.56 [-13.85, -1.27] | 0.98 [0.63, 1.53] | -8.33 [-14.62, -2.03] | 0.005 | 0.89 |
| Africa | South Africa (5) | LDC | 1975 | 0.95 [0.35, 1.56] | -7.16 [-15.33, 1.02] | 1.01 [0.57, 1.78] | -7.90 [-16.07, 0.27] | 0.015 | 0.89 |
| Asia | Bangladesh (5) | LDCL | 1975 | 0.72 [-0.00, 1.44] | -4.99 [-15.42, 5.43] | 0.82 [0.37, 1.81] | -6.44 [-16.87, 3.98] | 0.051 | 0.77 |
| Superlinear (RMA Exp. CI > 1) | | | | | | | | | |
| Europe | Italy (6) | MDR | 1975 | 1.26 [1.06, 1.47] | -12.39 [-15.26, -9.53] | 1.27 [1.09, 1.49] | -12.51 [-15.38, -9.64] | 0.000 | 0.99 |

**Supplemental 1. Scaling parameters of all countries over time.**

**Countries that changed from linear to sublinear:** Argentina, Australia, Bangladesh, Canada, China, Indonesia, South Korea, Spain, Turkey, United States, United Kingdom

**Countries that stayed linear:** Belgium, Colombia, Germany, Iran, Malaysia, Nigeria, Pakistan, Poland, Saudi Arabia, South Africa, Ukraine, Venezuela

**Countries that stayed sublinear:** Brazil, Egypt, France, India, Japan, Mexico, Russia, Taiwan, Uzbekistan

**Countries that changed from superlinear to linear:** Italy

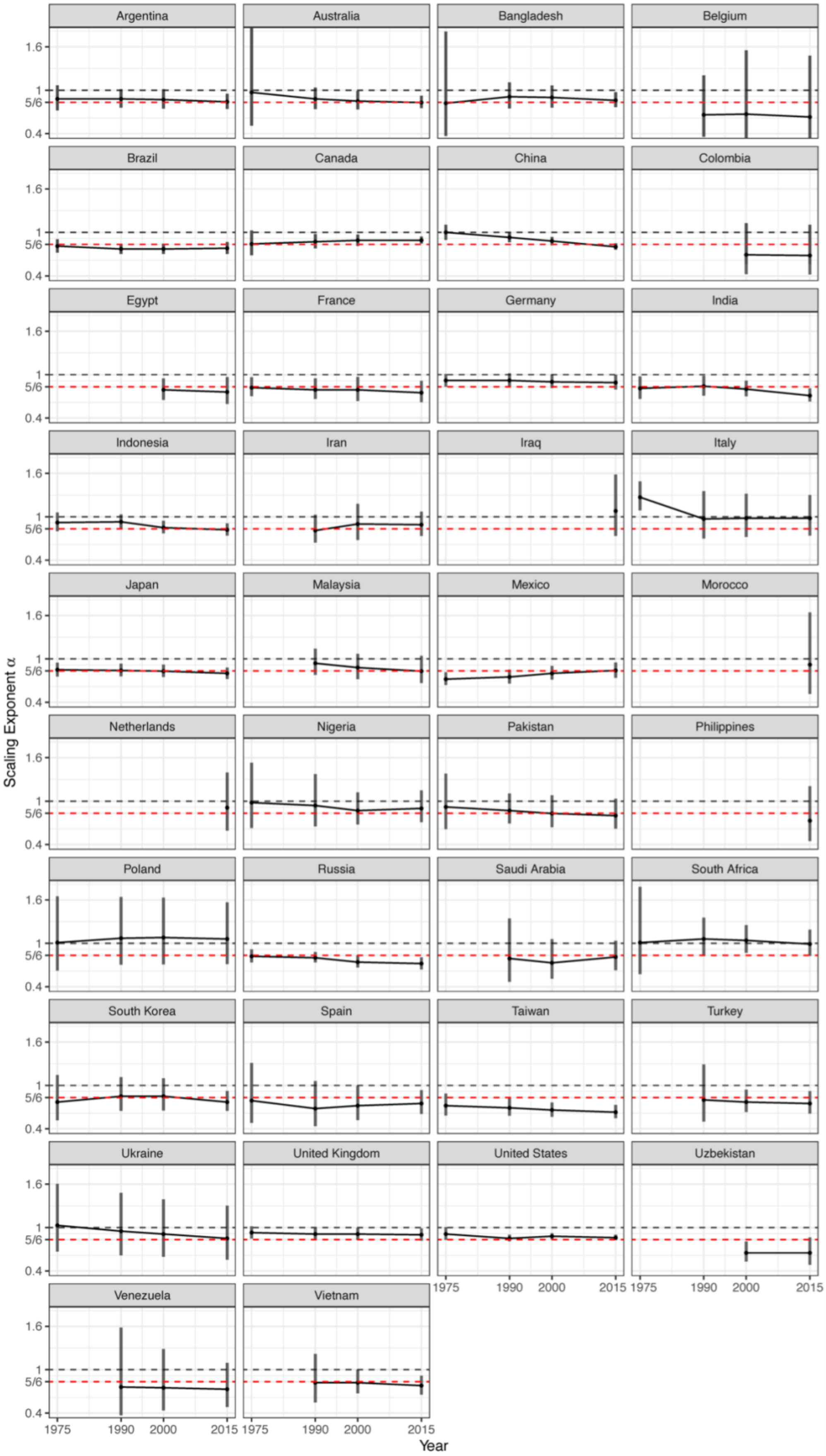